# Hybrid Quantum Applications Need Two Orchestrations in Superposition: A Software Architecture Perspective

Frank Leymann[0000-0002-9123-259X] and Johanna Barzen[0000-0001-8397-7973]

University of Stuttgart, IAAS, Universitätsstr. 38, 70569 Stuttgart, Germany
{firstname.lastname}@iaas.uni-stuttgart.de

**Abstract**: Quantum applications are most often hybrid, i.e. they are not only made of implementations of pure quantum algorithms but also of classical programs as well as workflows and topologies as key artifacts, and data they process. Since workflows and topologies are referred to as "orchestrations" in modern terminology (but with very different meaning), two orchestrations that go hand-in-hand are required to realize quantum applications. We motivate this by means of a non-trivial example, sketch these orchestration technologies and reveal the overall structure of non-trivial quantum applications as well as the implied architecture of a runtime environment for such applications.

**Keywords:** Quantum Software, Quantum Computing, NISQ, Software Engineering of Quantum Applications, Hybrid Quantum-Classical Applications, Runtime for Quantum Applications.

## 1. Introduction

Nowadays, quantum applications are in most cases hybrid, i.e. they encompass not only implementations of one or more quantum algorithms proper but require classical programs as well in order to produce their final result. This is most evident by the requirement for pre- and post-processing. For example, pre-processing generates circuits for state preparation within a classical environment and prepends these circuits to the quantum algorithm proper [LB20]. These state preparation circuits then create - when being executed - the quantum state that represents the input to be processed by the quantum algorithm. An example for post-processing is the correction of readout errors within a classical environment by applying an unfolding method to compute the (less) undisturbed result distribution from the disturbed measured distribution [LB20].

But even algorithms that are often considered as "proper quantum" algorithms are in fact hybrid. For example, the factorization algorithm of Shor consists of a quantum part that produces an output that must be post-processed by a classical program by means of analyzing continued fractions. I.e. the implementation of this quantum algorithm needs to integrate with a classical program to produce its final result.

In general, a hybrid quantum application (or quantum application for short) is not only made of implementations of quantum algorithms (called quantum programs from here on) and classical programs, but also of data to be processed, workflows, and topology models. The quantum programs may be written in a quantum assembler (like

OpenQASM or Quil) or a host programming language using quantum libraries (like QisKit or Cirq). The classical programs may be written in any language, run in any environment. Data may be provided by value or by reference (and then retrieved), have to be properly transformed etc. Workflows may be used for preparing data for further processing, for controlling the execution order of the quantum programs and classical programs, as well as passing data between these programs.

The structure of this paper is as follows: section 2 sketches a non-trivial quantum application that motivates the use of workflow technology as a major enabler for real-world quantum applications. Section 3 describes the need for the use of provisioning technology as the other major enabler for real-world quantum applications. The implied overall runtime environment for quantum applications is detailed in section 4. Section 5 sketches related work, and a conclusion and outlook is given in section 6.

## 2. Workflows: Orchestrating Control- and Data-Flow

In this section we briefly sketch the concept of a workflow model and a workflow instance, and describe a real-world sample hybrid quantum application.

### 2.1. Workflow in a Nutshell

Workflow technology is well-established since decades [LR00]. In a nutshell, it is a technology to specify the partial order of a collection of activities that have to be performed to achieve a composite goal. The partial order is based on control flow dependencies between the activities. Typically, the activities are represented as nodes in a directed graph (see Figure 1), and the control flow dependencies are the edges of the graph. Such an edge points from an activity to those activities that may have to be performed once the source activity finished successfully. Whether or not a target activity is actually performed is controlled by a Boolean condition associated with the corresponding edges. This condition is evaluated based on data that has been returned by already finished activities. This way, the set of activities performed by a workflow is highly dependent of the results of the activities. Consequently, the actual paths taken through the graph typically changes from execution to execution of the workflow. The directed graph representing the workflow is referred to as a *workflow model*, and an execution of such a workflow model is referred to as an *instance* of the model. Nowadays, workflow models are usually specified in BPMN [BPMN], which is a graphical language with an operational semantics describing how instances of a workflow graph are created. [LK10] gives an overview of several key languages for specifying workflow models.

### 2.2. A Sample Hybrid Quantum Application

Figure 1 shows a sample hybrid quantum application applying quantum machine learning in the domain of the humanities [BL19], [BL20], [B21]. The application will cluster a set of input data. For the this purpose, the input data will be prepared, next its features will be determined, and finally the clustering itself will be performed. The data preparation activities are all classical programs (indicated by the "hammer" icon associated with the activities): for example, the (categorical) data will be retrieved from a database, then the distance matrix of this categorical data are computed (which is based on the Wu-Palmer similarity - see [BL+21]) and so on. The feature engineering part of the workflow begins by computing the covariance matrix, transforms it into its Pauli representation and performs a sub-workflow (indicated by the "gear" icon) that computes the eigenvalues of the covariance matrix by especially using an activity that is executed on a quantum computer (indicated by the "atom" icon), and finally a sub-workflow made of classical programs performs a principal component analysis (as indicated, the eigenvalues are computed via the variational quantum eigensolver). After that clusters are determined which involves a sub-workflow that solves a maximum cut problem implemented by means of a quantum approximate optimization algorithm.

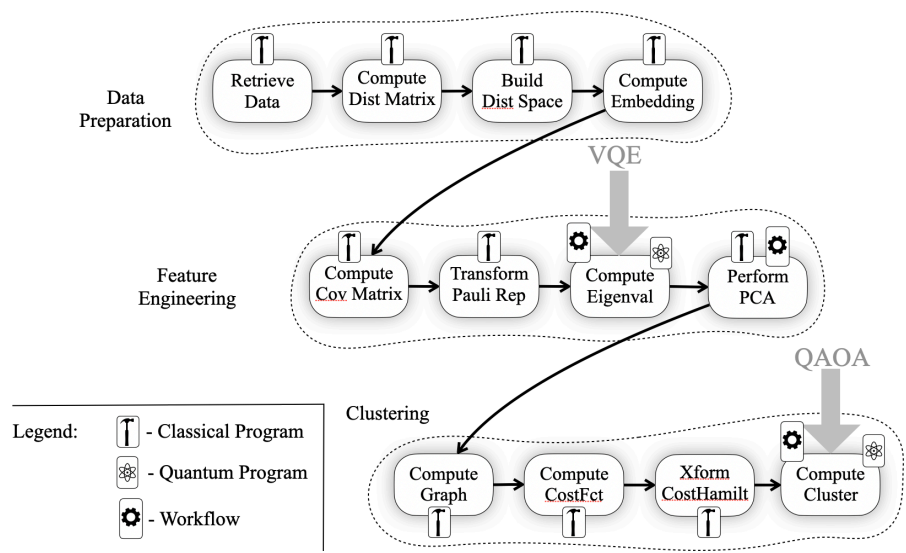

**Fig. 1**. The Workflow of the Sample Application.

### 2.3. Executing a Workflow

A workflow model will be executed by a *workflow engine*. For this purpose, the workflow engine navigates through the workflow model: for example, it determines the activities that are ready to be performed, collects their input data, starts (and controls) their execution (in parallel), retrieves their output, and determines once an activity completes successfully - based on the transition conditions of the edges leaving a completed activity - the activities to be executed next. Such a workflow in execution is referred to as an *instance* of the corresponding workflow model (see [LR00] for the details). Note, that especially in a cloud environment workflows are also called *orchestrations* - based on the mental model that a workflow "orchestrates" all the actions required to create a single whole from the executions of the individual activities.

Obviously, the workflow engine must know how an activity is implemented. Thus, the workflow model associates with each activity its implementation (e.g a classical program or a quantum program), or it specifies how an implementation can be discovered at runtime. Once an activity is determined to be ready for execution, its input data is gathered by the workflow engine, transformed into a format required by its implementation, and passed to the implementation. This implies that the workflow engine understands the invocation mechanism of the implementation (e.g. how to call a Java program, a Python program, how to serve a REST API, how to start another workflow, how to communicate via a message queue, and so on). Obviously, the different implementations of the different activities may not only be very heterogenous but especially highly distributed, and they may run in very different environments. Furthermore, some of the activity implementations may be long-running and return responses asynchronously at unforeseen times, which implies that the workflow engine must understand how to correlate incoming data with instances and running activities therein to detect their completion.

This in turn requires that the state of a workflow instance (consisting of the state of each of its activities, the input and output data of the activities and so on) must be made persistent by the workflow engine. Especially, this implies that the execution of a workflow is interruptible. Also, errors are detected by the workflow system and parts of a workflow instance can be automatically undone if such an error is detected. This requires that for activities to be undone a compensating activity is assigned which the workflow engine will invoke in case of an error. Activities may be grouped into corresponding units of work that have an all-or-nothing semantics. Thus, a workflow is recoverable.

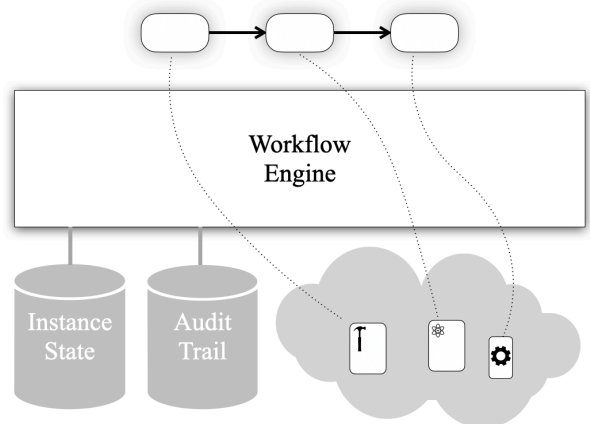

**Fig. 2**. Executing a Workflow.

Figure 2 shows a workflow engine and a workflow model navigated by this engine. The implementations corresponding to the activities are available to the workflow engine. When navigating the workflow model, the state information about the corresponding instance is stored in a database: this allows especially to monitor running instances. Once an instance is completed the information about the history of its execution (i.e. the steps performed, their duration, input/output data, reasons for taking a particular path etc) is moved to another database referred to as audit trail. The audit trail can be analyzed to improve workflows (making it faster, cheaper, correcting modeling errors etc) and enable reproducibility of results.

## 3. Provisioning: Orchestrating Topology Deployment

In this section we sketch how the environment required for executing a workflow is specified and automatically set up.

### 3.1. Topology of a Hybrid Quantum Application

Whenever an activity of a workflow is detected to be ready for execution, the corresponding activity implementation is invoked (see section 2.3). This assumes that the corresponding implementation is available in (or at least accessible from) the environment. Since an implementation (e.g. a Java program) typically has dependencies on other artifacts (like a JVM), these artifacts have to be available too - and in a transitive manner: only if all these artifacts are present and intertwined correctly, the activity implementation can be performed. All the necessary artifacts and their dependencies are described by a directed graph the nodes of which are the artifacts and the directed edges are the dependencies between the artifact. Such a graph is referred to as a *topology model*. [BB+12] describes a standardized language to specify topology models, [TO] is the described specification, and [BB+13], [BE16] sketches an associated open source eco-system.

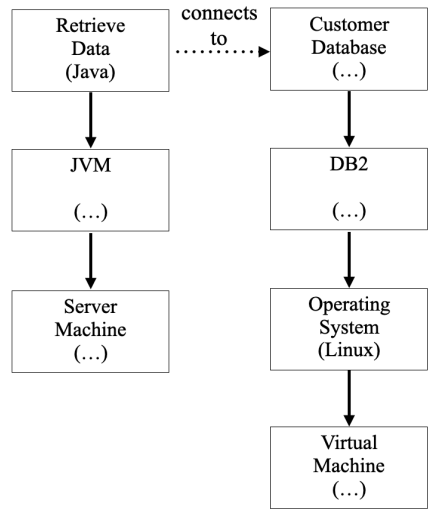

**Fig. 3**. A Sample Topology Model.

Figure 3 depicts an example of such a topology model. It shows the implementation of the "Retrieve Data" activity of the sample workflow from Figure 1 and its dependencies. This implementation is a Java program which obviously depends on a JVM. The JVM in turn requires a server machine it is installed on. Furthermore, the Java program needs to connect to the customer database to actually retrieve the data. The database is managed by a DB2 database system, which requires a Linux operating system, which in turn is hosted by a virtual machine.

### 3.2. Package of an Hybrid Quantum Application

A quantum application is delivered as a single entity referred to as *quantum application archive* (QAA). This archive is a self-contained package that encompasses all the artifacts needed to setup the execution environment required to perform the quantum application (see Figure 4). First, this package contains the topology model describing the artifacts and their dependencies. Next, all classical programs as well as all quantum programs making up the quantum application are included (or pointed to). Then, the workflow model orchestrating the execution of the activity implementations are in the package, as well as workflow models that are (re-)used as sub-workflows (like the "compute eigenvalue" sub-workflow realizing a variational quantum eigensolver in Figure 1). Finally, some quantum applications like a machine learning application for training a neural net may need special data; such data may be included in the package too. This way a quantum application becomes an entity like an app that can be stored somewhere (e.g. in some sort of a quantum app store), advertised, bought and so on.

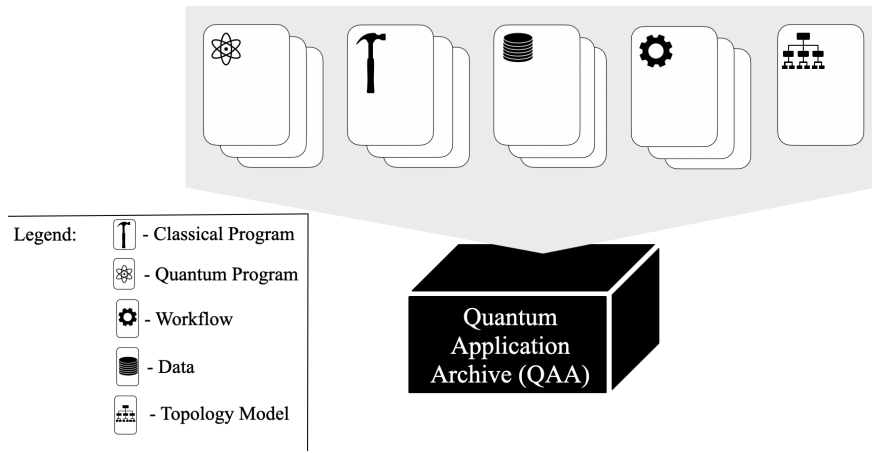

**Fig. 4**. Application Package

### 3.3. Provisioning an Execution Environment

Before the corresponding quantum application can be executed, its required execution environment must be setup. For this purpose, the topology model of the execution environment is interpreted by a provisioning engine. In a nutshell, the *provisioning engine* interprets the topology model "from the bottom to the top", i.e. from the leafs of the topology model in reverse direction of the edges of the corresponding topology graph. For each node visited, the corresponding artifact is installed.

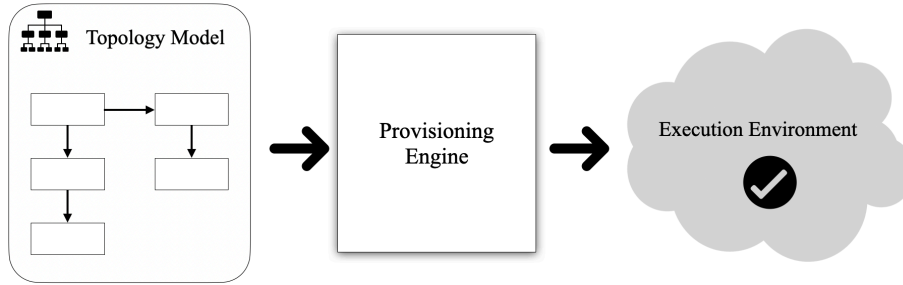

**Fig. 5**. Provisioning of an Application Package.

For example, in Figure 3 the server machine is allocated, the JVM is installed on this machine, and the Java application is deployed in the JVM. In parallel, the database is installed. Finally, the connection between the Java program and the database system is established such that at runtime the Java program can access the database and retrieve data: see [BB+12] for more details. Figure 5 depicts that the provisioning engine interprets a topology model and provisions the corresponding execution environment. Note that in analogy to workflows, the interpretation of topology models for provisioning an execution environment is called *orchestration* too - the mental model is again that the provisioning engine "orchestrates" all the actions required to create an execution environment as a whole by deploying individual artifacts.

## 4. The Quantum-Classical Environment

The technologies and concepts that have been described in this paper before imply architectural components of an environment for executing quantum applications. This section outlines these implications.

### 4.1. High-Level Architecture

Every quantum application includes quantum programs, thus, the runtime environment of quantum applications has to encompass one or more quantum processing units (QPU). Since QPUs are nowadays made available by means of cloud access [LB+20], it is only natural to assume that the classical programs of a quantum application are running in a cloud environment too (see Figure 6). Note that the latter is without loss of generality because workflow engines can invoke implementations that run in non-cloud environments too. Having said that, the runtime environment obviously must contain a workflow engine. To setup the environment for running the implementations of the activities of the workflow, a provisioning engine must be available too. Both, the workflow engine as well as the provisioning engine are also hosted in the cloud.

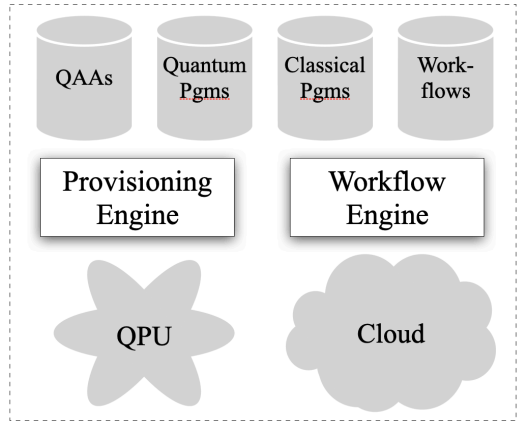

**Fig. 6**. Ingredients of a Hybrid Quantum-Classical Runtime Environment

Consequently, all the artifacts that make up a quantum application have to be accessible in the cloud. First of all, the quantum application archives are needed by the provisioning engine to set up the environment of the implementations of the activities of the workflows of the quantum application. The provisioning engine will process the QAA of a quantum application by orchestrating the deployment of the topology included in the QAA. During this processing, the quantum programs (or circuits) and classical programs will be installed and their prerequisites will be made available too: i.e. the runtime environment for the quantum application is set up. Next, the workflow engine will instantiate the workflow model representing the quantum application, i.e. the workflow models have to be part of the overall environment too.

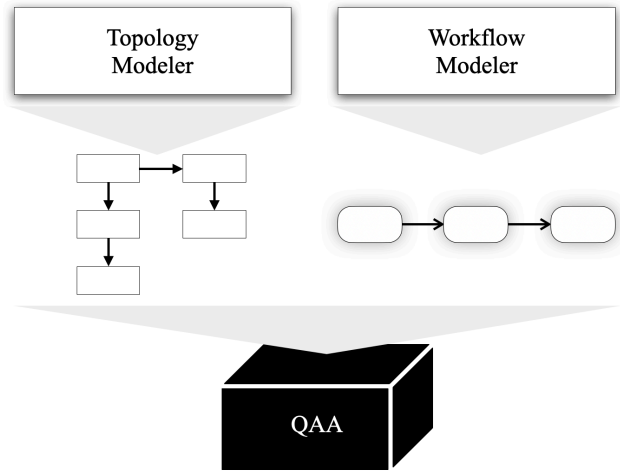

**Fig. 7**. Ingredients for a Hybrid Quantum-Classical Modeling Environments

In addition to the runtime environment, an environment for specifying the artifacts that make up a quantum application is needed. Beside tools well-known to quantum programers like a circuit designer, a modeling tool for topologies as well as a modeling tool for workflows are required. Figure 7 depicts these components of the modeling environment and that their output can be packaged into a QAA.

### 4.2. Running a Hybrid Quantum Application

Figure 8 depicts how the execution of a hybrid quantum application is kicked-off. Basically, a corresponding RUN message specifying the name of the workflow $\Omega$ and the initial parameter values $p_1,\ldots,p_k$ to be passed to the newly created workflow instance is put into a queue. This queue is the entry into the hybrid quantum-classical environment.

A dedicated component (called queue controller, for example) monitors the queue, analyzes the messages, and forwards it to the responsible components for further processing. In our context, the queue controller understands that the message solicits to instantiate the workflow model $\Omega$ and, thus, passes a corresponding request to the workflow engine. The workflow engine will fetch the workflow model $\Omega$ and will create a new instance passing the parameter values $p_1,\ldots,p_k$ to it as input.

This assumes that the environment needed by the workflow for its execution has already been deployed. To reduce costs incurred by cloud resources for workflows that are only rarely performed, the environment may be created for each execution of a workflow and may be deconstructed once the workflow finishes. If a workflow model is instantiated and run very often, deconstructing the corresponding environment and provisioning it over and over again may turn out to be a significant overhead and should be avoided. However, a corresponding component like a resource manager may be in charge of such decisions.

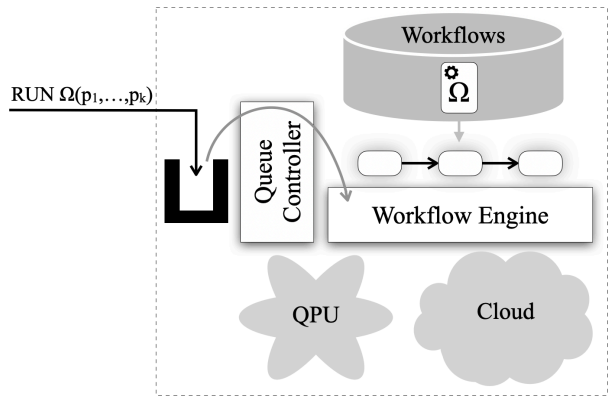

**Fig. 8**. Starting the Execution of a Hybrid Quantum Application

The above assumes that the quantum application archive has been unpacked before and that its encompassed artifacts are accessible to the provisioning engine, e.g. stored in the environment. If this is not intended (e.g. in order to avoid storage costs), another possibility is a variant of the RUN message that allows to put a complete quantum application archive for processing into the queue (Figure 9).

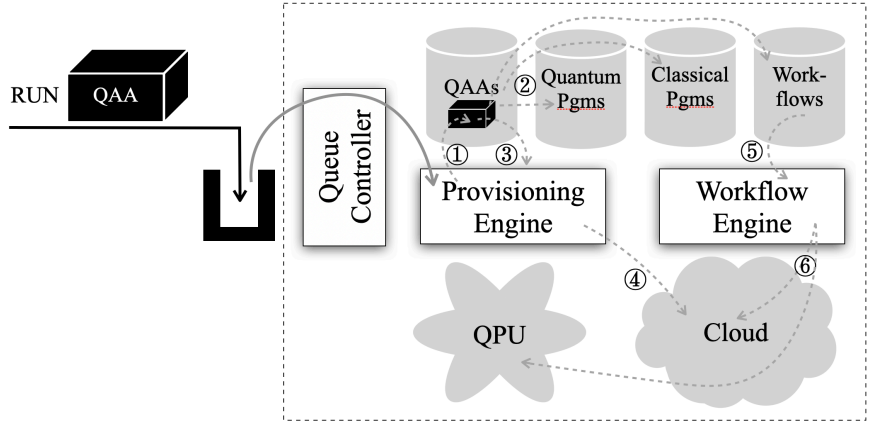

**Fig. 9**. Passing an Application Package for Immediate Execution

In this case, the queue controller will request the provisioning engine to deploy the quantum application archive and next request the workflow engine to run the contained hybrid quantum application - from outside, this is perceived as a single step [VK+13]: the QAA will be stored (temporarily) to be able to retry deploying the QAA in case of errors (step ①). Next (step ②), the archive is unpacked: the quantum programs, the classical programs, and the workflows are stored such that they are individually accessible. In step ③, the provisioning engine will interpret the topology model of the QAA and determine all actions needed to setup the environment required by the hybrid quantum application. Once all these actions are performed, the environment needed by the workflow for its execution is deployed (step ④). Now, the main workflow model of the hybrid quantum application will be instantiated (step ⑤) and executed. During its execution the workflow engine will invoke classical programs deployed in the cloud and will kickoff the execution of quantum programs on a QPU (step ⑥).

### 4.3. Hybrid Quantum Applications and Two Orchestrations in Superposition

As a consequence, two different kinds of orchestrations are required to perform a hybrid quantum application: one orchestration of the control- and data-flow between the activities of the quantum application, and another orchestration of the deployment of the topology of the environment required by workflow to be executed (Figure 10). The figure also shows, that these two kinds of orchestrations are intertwined [WBr+20], they are in a loose sense in superposition: the orchestration performed by the workflow engine (the x-axes) goes hand in hand with the orchestration performed by the provisioning engine (the y-axes). The workflow engine instantiates the workflow model and invokes activity A, which requires that the implementation ① of activity A is properly setup; and this setup is based on the corresponding fragment of the topology of the overall environment required by the workflow as indicated in the figure. Once activity A completed successfully, the implementation ② of activity B is started, which assumes a proper setup corresponding to another fragment of the overall topology. Finally, activity C's implementation ③ is kicked-off and the setup of this implementation is specified by yet another topology fragment.

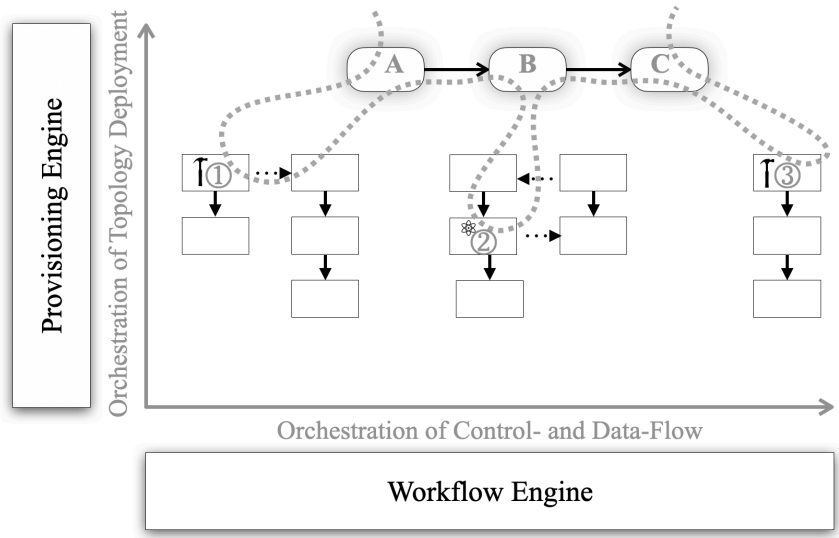

**Fig. 10**. Executing a Quantum Application Requires Two Orchestrations in Superposition.

### 4.4. Optimization Potentials: Example

The graph-based nature and corresponding operational semantics of (many) orchestration languages support their formal analysis for predictions or improvements, for example. This offers opportunities for a better support of quantum programs that iteratively access a quantum computer like implementations of variational quantum algorithms that are often used today with NISQ devices. The iterative nature of such an algorithm, i.e. the invocation of quantum program in a loop that typically also contains classical programs, can easily be detected in a workflow model.

In such a situation, the hybrid quantum-classical environment may reserve sole access to the quantum computer for the corresponding quantum application and provide direct access to the quantum computer without having to submit requests via the queue (see section 4.2). This will reduce the latency of the quantum application significantly.

## 5. Related Work

A workflow language can be perceived as a parallel, long-running, interruptible, persistent and recoverable programming language (section 2.3). The instantiation of a workflow model by a workflow engine ensures these properties: the workflow engine acts like a virtual machine properly interpreting the workflow language, just like a Java Virtual Machine interprets the Java language. Attempting to use a traditional programming language to realize "workflows" will thus fail. Consequently, domains in which workflows play a central role either use an existing *workflow system* (i.e. a workflow engine and a matching modeling tool), or develop a separate workflow system that is targeted to the particular application domain. Note, that the latter is a huge endeavor. The use of an existing workflow system has the advantage that it is mature, proven, robust and so on.

The latter situation occurred, for example, in the domain of eScience [HT09] where a plethora of workflow systems (called scientific workflow systems) has been developed [LV15]. As a result, workflows are hard to reuse cross domains of eScience or cross scientific workflow systems because of a lack of standardization. Also, these workflow systems typically are not abreast in terms of maturity etc with conventional workflow systems. And it turned out, that conventional workflow systems in fact can be used for scientific workflows either without any modifications or with a few proper extensions [GS11].

In the quantum computing domain, history seems to repeat. First workflow systems dedicated to quantum computing appeared in the scientific domain like Nexus [K16], but also product offerings specialized for quantum computing like Zapata Orquestra [Z21] are made available. In this context too, it turned out, that conventional workflow systems can be used for quantum workflows with only a very few extensions [WeB+20] guaranteeing to benefit from the maturity and robustness of conventional workflow systems. And by using standardized workflow languages like BPMN [BPMN], reuse of workflows across workflow engines is simplified.

Note, that nowadays conventional workflow systems support one of two standard workflow languages: BPEL [BPEL] or BPMN [BPMN], while BPMN is becoming the dominant language. An overview of related standards and the concept behind the languages can be found in [LK10]. Using one of these two languages eases the reuse of workflows across supporting workflow engines. Extensions of BPMN for quantum computing have been proposed in [WeB+20] and have been prototypically implemented based on the open source BPMN workflow system Camunda [Ceng21].

[L12] introduced the concept of self-contained application archives especially for the purpose of understandability and reproducibility of scientific in-siloco experiments. This concept has been realized in [WBr+20] introducing self-contained archives consisting of workflows and all of its dependencies, both, in terms of modeling as well as automatically provisioning the complete environment required by a workflow for its execution.

In analogy to workflow technology, a plethora of provisioning or deployment technologies, respectively, like Kubernetes, Puppet, Ansible etc have been proposed and are used in practice today. TOSCA (see [TO], [BB+12]) is a standardized language for modeling topologies and their operational semantics. It contains a subset that can be mapped to the former technologies [WuB20]. An open source ecosystem for modeling topologies based on TOSCA and executing their deployment as well as support of the subset mappable to the former technologies has been provided (see [BB+13], [BE16]). Finally, extensions of TOSCA have been developed that support the modeling and provisioning of quantum applications [WiB+20].

## 6. Conclusion and Outlook

Most quantum applications are hybrid, i.e. they consist of both, quantum programs as well as classical programs. This implies that the control- and data-flow between the corresponding artifacts as well as the proper deployment of the implementations of these artifacts themselves need to be orchestrated, and both orchestrations must be inter-twinned. We elucidated this by means of a real-world use case. In order to treat quantum applications as a self-contained entity, we introduced the quantum application archive (QAA) that collects all the artifacts of a quantum application as well as all the required information for their processing in a single package. The use of proven workflow technology for orchestration of the flow between the artifacts of a quantum application as well as the use of proven provisioning technology for orchestration of the topology of a quantum application has been argued for. We sketched the high-level architecture of a runtime environment for quantum applications and especially revealed the role of a workflow engine and a provisioning engine in such a runtime; also the need for a workflow modeling tool and a topology modeling tool as components of a build time environment for quantum applications have been mentioned.

The sketched architecture is currently verified by an initial prototype based on the Camunda workflow system [Ceng21] and its associated modeling tool [Cmod21] for running and modeling the orchestrations of the control- and data-flow of a quantum application. As workflow language, BPMN has been chosen (and had to be extended to be able to interact with quantum computers [WeB+20]). Modeling of topologies is done via the modeling tool Winery [W21], and the orchestration of the deployment of the topology models is performed via OpenTOSCA [Oto21]. As topology language, TOSCA [TO] has been chosen (again with proper extensions [WiB+20]). Further extensions of this prototype are worked on, especially the integration between the queue controller on one side, and the workflow engine and provisioning engine on the other side is still open. The same is true for the indicated optimization to reserve a quantum computer to a quantum application in iterations.

**Acknowledgements**

We are very grateful for the discussions about selective subjects of this paper with our colleagues Benjamin Weder, Karoline Wild and Michael Zimmermann as well as their work on the prototype, which is also significantly supported by Daniel Fink and Felix Truger.

This work was partially funded by the BMWi project PlanQK (01MK20005N), as well as the project SEQUOIA funded by the Baden-Wuerttemberg Ministry of the Economy, Labour and Housing.